\documentclass{article}
\usepackage{spconf} %
\usepackage{graphicx}
\usepackage{amssymb}
\usepackage{amsmath}
\usepackage{hyperref}
\usepackage[utf8]{inputenc}
\usepackage[table,xcdraw]{xcolor}

\newcommand{\B}{\mathbf{B}}
\newcommand{\I}{\mathbf{I}}
\newcommand{\K}{\mathbf{K}}
\newcommand{\R}{\mathbf{R}}

\newcommand{\X}{\mathbf{X}}

\newcommand{\bK}{\mathbf{K}}
\newcommand{\bc}{\mathbf{c}}
\newcommand{\bk}{\mathbf{k}}

\newcommand{\f}{\mathbf{f}}

\newcommand{\x}{\mathbf{x}}
\newcommand{\y}{\mathbf{y}}

\newcommand{\bbc}{\bar{\bc}}
\newcommand{\bby}{\bar{\y}}
\newcommand{\sK}{\boldsymbol{\mathcal{K}}}

\newcommand{\Real}{{\mathbb R}}

\newcommand{\Normal}[0]{\mathcal{N}}

\hypersetup{
    unicode=false,          %
    pdftoolbar=true,        %
    pdfmenubar=true,        %
    pdffitwindow=false,     %
    pdfstartview={FitH},    %
    pdftitle={Gapfilling},    %
    pdfauthor={Camps-Valls},     %
    pdfsubject={Gapfilling},   %
    pdfcreator={Camps-Valls},   %
    pdfproducer={Camps-Valls}, %
    pdfkeywords={keyword1} {key2} {key3}, %
    pdfnewwindow=true,      %
    colorlinks=true,       %
    linkcolor=blue,          %
    citecolor=blue,        %
    filecolor=blue,      %
    urlcolor=blue           %
}

\begin{document}

\title{Gap filling of biophysical parameter time series\\
with multi-output Gaussian processes}
\makeatletter
\def\@name{
\emph{Anna Mateo-Sanchis$^1$, Jordi Muñoz-Marí$^1$, Manuel Campos-Taberner$^2$,} \\
\emph{Javier Garc\'ia-Haro$^2$, Gustau Camps-Valls$^1$} \\
\thanks{This paper has been partially supported by the European Research Council under Consolidator Grant SEDAL ERC-2014-CoG 647423, and MINECO/ERDF under Grant CICYT TIN2015-64210-R.
Published in 2018 IEEE International Geoscience and Remote Sensing Symposium.}}
\makeatother
\address{$^1$Image Processing Laboratory (IPL). Universitat de Val\`encia, Spain.\\
$^2$Dep. Física de la Terra i Termodinàmica, Universitat de Val\`encia, Spain.}

\maketitle

\begin{abstract}
In this work we evaluate multi-output (MO) Gaussian Process (GP) models based on the linear model of coregionalization (LMC) for estimation of biophysical parameter variables under a gap filling setup. In particular, we focus on LAI and fAPAR over rice areas. We show how this problem cannot be solved with standard single-output (SO) GP models, and how the proposed MO-GP models are able to successfully predict these variables even in high missing data regimes, by implicitly performing an across-domain information transfer.
\end{abstract}

\section{Introduction}

Monitoring vegetation from space is of paramount importance.
The Leaf area index (LAI) and the Fraction of Absorbed Photosynthetically Active Radiation (fAPAR) are among the most important essential climate variables (ECVs)~\cite{GCOS11} for land and vegetation monitoring. LAI is a key bio-physical parameter which represents half of the total leaf area per unit of ground area~\cite{Chen92}, while fAPAR accounts for the light absorption across an integrated plant canopy. Both variables are used as indicators of the state and evolution of the vegetation cover.

LAI and fAPAR have been extensively used in many agricultural and remote sensing studies~\cite{Myneni97,Campos16}, %
and are key in climate models~\cite{Sellers96}. Both products are assimilated into physical models to describe vegetation processes, such transpiration and photosynthesis, as well as machine learning models to estimate carbon, energy and heat fluxes~\cite{CampsValls15igarss}.

It goes without saying that the precise estimation and modeling of the evolution through time of these parameters have deep societal, economical and environmental implications.
Unfortunately, LAI and fAPAR time-series often exhibit high rates of missing data due to the presence of clouds and snow. Gaps within the time-series reduce their usefulness for modeling and monitoring environmental phenomena~\cite{Weiss14}. In this context, gap filling and interpolation with statistical and process-based approaches has become a successful choice in the last decade.
Fusion techniques and regression trees were used in image mosaics reconstruction without clouds \cite{Pohl99, Homer97}. Kriging and co-kriging techniques were applied for spatial interpolation of missing data in Landsat images \cite{Zhang2007,Zhang09}. %
Recently,~\cite{Kandasamy13} compared  gap filling and interpolation methods on MODIS LAI products concluding that, in general, temporal smoothing techniques performed better than the rest, especially in high ($>20\%$)  missing data regimes. %

In this work, we aim to explore advanced machine learning approaches for interpolation. In particular we will focus on Gaussian processes (GPs) for regression~\cite{Rasmussen06}, which has recently provided very good results for model inversion and bio-physical parameter estimation~\cite{CampsValls16grsm}. Noting the tight association between LAI and fAPAR, we aim to model them together using multi-output Gaussian Processes (MO-GP). We do this under the linear model of coregionalization (LMC) framework~\cite{Alvarez11}. These models take into account the relationships among output variables learning a cross-domain kernel function able to {\em transfer} information between time series. The learned relations are exploited to do inferences on regions where no training samples (gaps) are available for one of the two variables. In contrast, the standard procedure of using individual GPs for each variable cannot make reliable predictions on areas with missing training samples.

The paper is organized as follows. Section~\ref{sec:methods} briefly reviews the standard Gaussian Processes formulation, and introduces the linear model of co-regionalization for multi-output regression problems. Section~\ref{sec:data} describes the datasets used and the experiments and results. Section~\ref{sec:conclusions} draws final conclusions and outlines future work.
\vspace{-0.35cm}
\section{Multiple-output Gaussian Processes} \label{sec:methods}

\vspace{-0.4cm}
This section first fixes the notation and reviews the standard GP formulation, and then introduces the linear model for co-regionalization in GPs to tackle multiple-output problems.

\subsection{Gaussian Process Regression}
GPs are state-of-the-art statistical methods for regression and function approximation, and have been used with great success in biophysical variable retrieval by following  statistical and hybrid approaches~\cite{Verrelst20121832}. %
We start assuming we are given a set of $n$ pairs of measurements, %
$\{\x_i,y_i\}_{i=1}^n$, perturbed by an additive independent noise.
We consider the following model,
\begin{equation}\label{GLR}
y_i = f(\x_i) + e_i,~~~e_i \sim\Normal(0,\sigma_n^2),
\end{equation}
where $f(\x)$ is an unknown latent function, $\x\in\Real^d$, and $\sigma_n^2$ represents the noise variance. Defining $\y=[y_1, \ldots ,y_n]^\intercal$ and $\mathbf{f}=[f(\x_1),\ldots, f(\x_n)]^\intercal$, the conditional distribution of $\y$ given $\mathbf{f}$ becomes $p(\y|\mathbf{f})=\mathcal{N}(\mathbf{f}, \sigma_n^2\I)$, where $\I$ is the $n\times n$ identity matrix. It is assumed that $\mathbf{f}$ follows a $n$-dimensional Gaussian distribution $\mathbf{f}\sim\mathcal{N}(\boldsymbol{0},\bK)$. %
The covariance matrix ${\bf K}$ of this distribution is determined by a squared exponential (SE) kernel function with entries $\bK_{ij}=k(\x_i,\x_j)=\exp(-\|\x_i-\x_j\|^2/(2\sigma^2))$, encoding the similarity between input points \cite{Rasmussen06}. %
In order to make a new prediction $y_\ast$ given an input $x_\ast$ we obtain the joint distribution over the training and test points,
\begin{equation}
\begin{bmatrix}
  \y \\ y_\ast
\end{bmatrix}
\sim \mathcal{N} \left( \mathbf{0},
\begin{bmatrix}
  \boldsymbol{C_n} & \bk_\ast^\intercal \\ \bk_\ast & c_\ast
\end{bmatrix} \right),  \nonumber
\end{equation}
where $\boldsymbol{C_n} = \bK + \sigma_n^2\I $, $\bk_{*} = [k(\x_*,\x_1), \ldots, k(\x_*,\x_n)]^\intercal$ is an $n\times 1$ vector and $c_\ast = k(\x_*,\x_\ast) + \sigma_n^2$. Using the standard Bayesian framework we obtain the distribution over $y_\ast$ conditioned on the training data, which is a normal distribution with predictive mean and variance given by
\begin{align} \label{eq:gppred}
  \begin{aligned}
   \mu_{\text{GP}} (\x_\ast) &= \bk_{*}^\intercal (\bK + \sigma_n^2\I_n)^{-1}\y, \\
   \sigma^2_{\text{GP}} (\x_\ast) &= c_\ast - \bk_{*}^\intercal (\bK + \sigma_n^2\I_n)^{-1} \bk_{*}.
  \end{aligned}
\end{align}
One of the most interesting things about GPs is that they yield not only predictions $\mu_{\text{GP}\ast}$ for test data, but also the uncertainty of the mean prediction, $\sigma_{\text{GP}\ast}$. Model hyperparameters $\boldsymbol{\theta}=[\sigma, \sigma_n]$ determine, respectively, the width of the SE kernel function and the noise on the observations, and they are usually obtained by maximizing the marginal likelihood.

\subsection{Coregionalization for GPs}
One the problems with the standard GPR formulation %
is that it applies only to scalar functions, i.e., we can predict only one variable. %
A straightforward strategy to deal with several target variables is to develop as may individual GP models as variables. While generally good performance is attained in practice, the approach has a clear shortcoming: %
the obtained models are independent and they do not take into account the relationships between the output variables. %
In order to handle this problem, we propose a multi-output GP model based on the \emph{linear model of coregionalization} (LMC)~\cite{Alvarez11}, also known as \emph{co-kriging} in the field of geostatistics~\cite{Journel78}.

In the {multi-output GP} model we have a vector function, $\f:\mathcal{X}\rightarrow R^D$, where $D$ is the number of outputs. Given a reproducing kernel, defined as a positive definite symmetric function {$\K:\mathcal{X}\times\mathcal{X}\rightarrow\Real^{N\times N}$, where $N$ is the number of samples of each output. In the following, and in order to simplify the equations that follows, we assume that all outputs have the same number of training samples, $N$. The formulation where each output has a different number of sources can be straightforwardly obtained}. We can express $\f(\x)$ as
\begin{equation}
\f(x) = \sum_{i=1}^N\K(\x_i,\x)\bc_i,  %
\end{equation}
for some coefficients {$\bc_i\in\R^N$. These coefficients can be obtained by solving the linear system, obtaining
\begin{equation}
\bbc = (\sK(\X,\X) + \lambda N\I)^{-1}\bby,
\end{equation}
where $\bbc,\bby$} are $ND$ vectors obtained by concatenating the coefficients and outputs, respectively, and {$\sK(\X,\X)$} is an $ND\times ND$ matrix with entries $(\K(\x_i,\x_j))_{d,d'}$ for $i,j=1,\ldots,N$ and $d,d'=1,\ldots,D$. The blocks of this matrix are $(\K(\X_i,\X_j))_{i,j}$ $N\times N$ matrices. %
Predictions are given by
{
\begin{equation}
\f(\x_*) = \sK_{\x_*}^\top\bbc,
\end{equation}
with $\sK_{x_*}\in\R^{D\times ND}$} composed by blocks $(\K(\x_*,\x_j))_{d,d'}$.

When the training kernel matrix {$\sK(\X,\X)$} is block diagonal, that is, $(\K(\X_i,\X_j))_{i,j} = \boldsymbol{0}$ for all $i\neq j$, then each output is independent of the others, and we have individual GP models. The non-diagonal matrices establish relationships among the outputs.

In the linear model of coregionalization (LMC) each output is expressed as a linear combination of independent latent functions~\cite{Journel78},
\begin{equation} \label{eq:lmc}
f_d(\x) = \sum_{q=1}^Q a_{d,q}u_q(\x),
\end{equation}
where $a_{d,q}$ are scalar coefficients, and $u_q(\x)$ are latent functions with zero mean and covariance $k_q(\x,\x')$. It can be shown~\cite{Alvarez11} that the full covariance (matrix) of this model can be expressed as
{
\begin{equation} \label{eq:sos-lmc}
\sK(\X,\X) = \sum_{q=1}^Q \B_q \otimes k_q(\X,\X),
\end{equation}
}
where $\otimes$ is the Kronecker product. Here each $\B_q\in R^{D\times D}$ is a positive definite  matrix known as \emph{coregionalization matrix},%
and it encodes the relationships between outputs.%

\section{Data collection and experimental results} \label{sec:data}

\subsection{Data collection}

In this study, we focus on the LAI and fAPAR biophysical variables from the Moderate Resolution Imaging Spectroradiometer (MODIS) products. In particular, both variables were obtained from the Collection-5 MOD15A2 1-kilometer resolution product on a sinusoidal grid. The temporal resolution of LAI/fAPAR is eight days based on a daily composition, which allows to obtain 46 estimates every year. Specifically, we focused on inter-annual variability of rice areas located in the València rice district (Mediterranean coast in Iberian peninsula) from 2003 to 2014. Typical rice plant phenology exhibits LAI values varying from zero (seeding) up to 6-7 (flowering), whereas fAPAR ranges from 0 to 1. %

\subsection{Experimental setup}

For the experiments, we simulate an scenario where we have missing data at the two ends of both time series. For LAI, the missing data is during the first years while for fAPAR is during the last years. Our goal is to obtain a model able to predict the missing parts in one variable using the information available from the other. %
The experiment was repeated with different number of missing years. %

We compared two GP models. First, we trained individual, single-output (SO-GP) models for each variable. The rational behind this first experiment is to assess how well SO models predict the variables in the regions with gaps, i.e., those without training samples. In the second experiment, we used MO-GP models. In this case, these models should be able to make better predictions on `gap' regions using the information inferred from the second, correlated variable.

\subsection{Qualitative evaluation}

\begin{figure}[t!]   %
\centerline{\includegraphics[width=11cm,height=7cm]{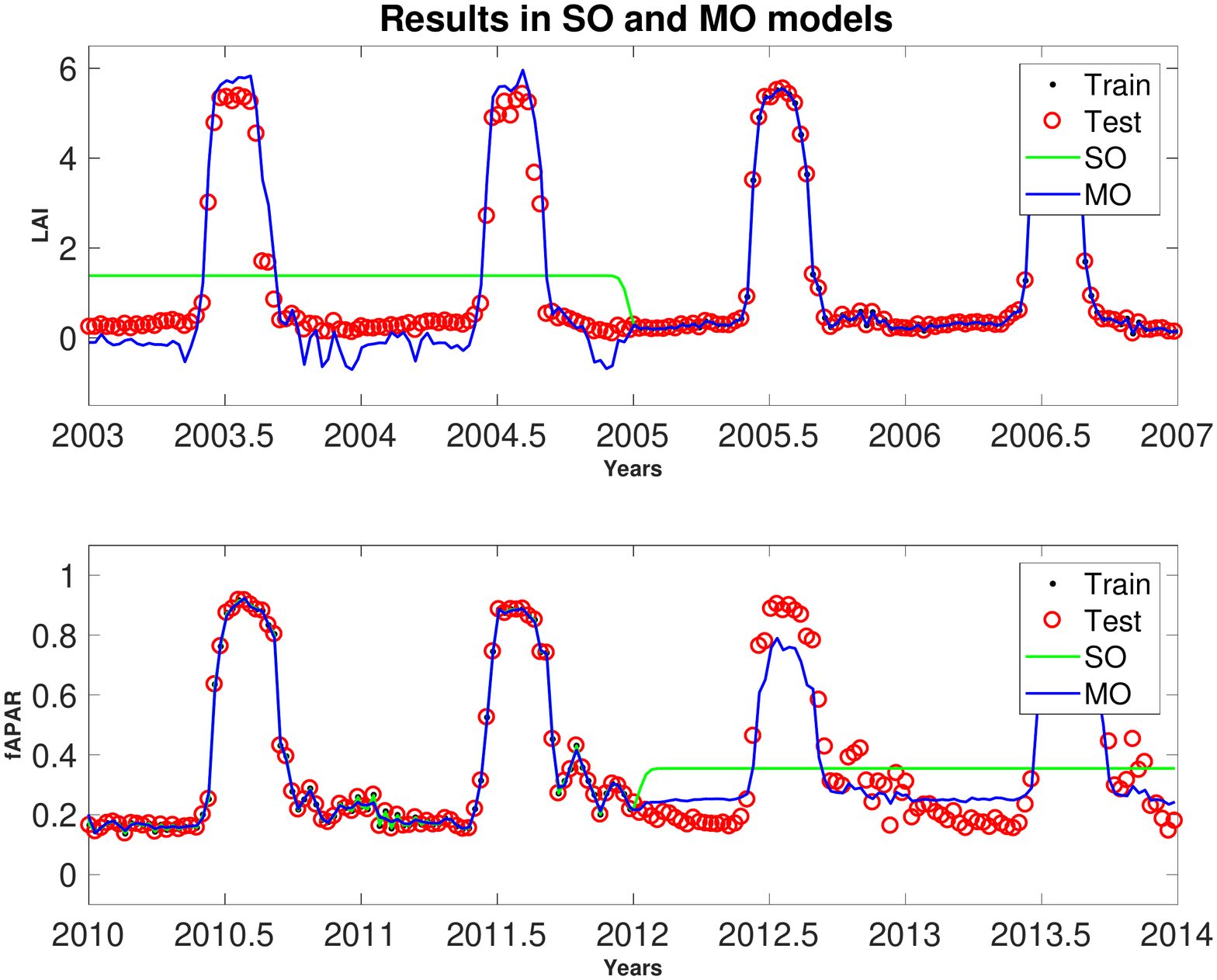}}
\vspace{-0.35cm}
	\caption{Individual (SO, green) and multi-output (MO, blue) predictions for LAI (top) and fAPAR (bottom).}
    \label{fig:MOSOresults}
\end{figure}

Figure~\ref{fig:MOSOresults} shows the predictions obtained with SO and MO GPs. For improved visualization, we focus in the missing years period only. %
The SO models (green line) are unable to make predictions where training data are missing, while the MO GPs  (blue line) can predict efficiently the variables in those regions, performing a sort of transfer learning across variables. %

\subsection{Performance in missing data regimes}

Table~\ref{tab:finalresults} shows the determination coefficient $R^2$ and the root mean square error (RMSE) in the test set for both models and different number of missing years (gaps). SO-GP models achieves an $R^2$ lower than 0.90, whereas MO-GP obtain much better results, close to 1. When comparing the RMSE, SO results are three times bigger than MO for both variables. %

\begin{table}[h!]
\centering
\small
\caption{Results in terms of R$^2$ and RMSE for different gaps.}
\begin{tabular}{|lc|ll|ll|}
\hline
\rowcolor[gray]{.60}
& & \multicolumn{2}{c}{LAI} & \multicolumn{2}{|c}{fAPAR} \\ \hline
\rowcolor[gray]{.90}
{\bf Model} & {\bf $\#$ miss. years} &{\bf R$^2$} & {\bf RMSE}& {\bf R$^2$} & {\bf RMSE}\\
\hline
SO-GP & 1 & 0.914 & 0.553 & 0.894 & 0.089 \\
MO-GP & 1 & 0.992 & 0.174 & 0.986 & 0.033 \\
\hline
SO-GP & 2 & 0.830 & 0.779 & 0.827 & 0.113 \\
MO-GP & 2 & 0.987 & 0.231 & 0.978 & 0.042 \\
\hline
SO-GP & 3 & 0.737 & 0.968 & 0.744 & 0.138 \\
MO-GP & 3 & 0.982 & 0.289 & 0.971 & 0.049 \\
\hline
SO-GP & 4 & 0.652 & 1.113 & 0.647 & 0.162 \\
MO-GP & 4 & 0.978 & 0.332 & 0.964 & 0.056 \\
\hline
SO-GP & 5 & 0.559 & 1.253 & 0.549 & 0.183 \\
MO-GP & 5 & 0.966 & 0.387 & 0.955 & 0.063 \\
\hline
\end{tabular}
\label{tab:finalresults}
\end{table}

\begin{figure}[t!] %
	\hspace{-1cm}
	\includegraphics[width= 10cm,height=4cm]{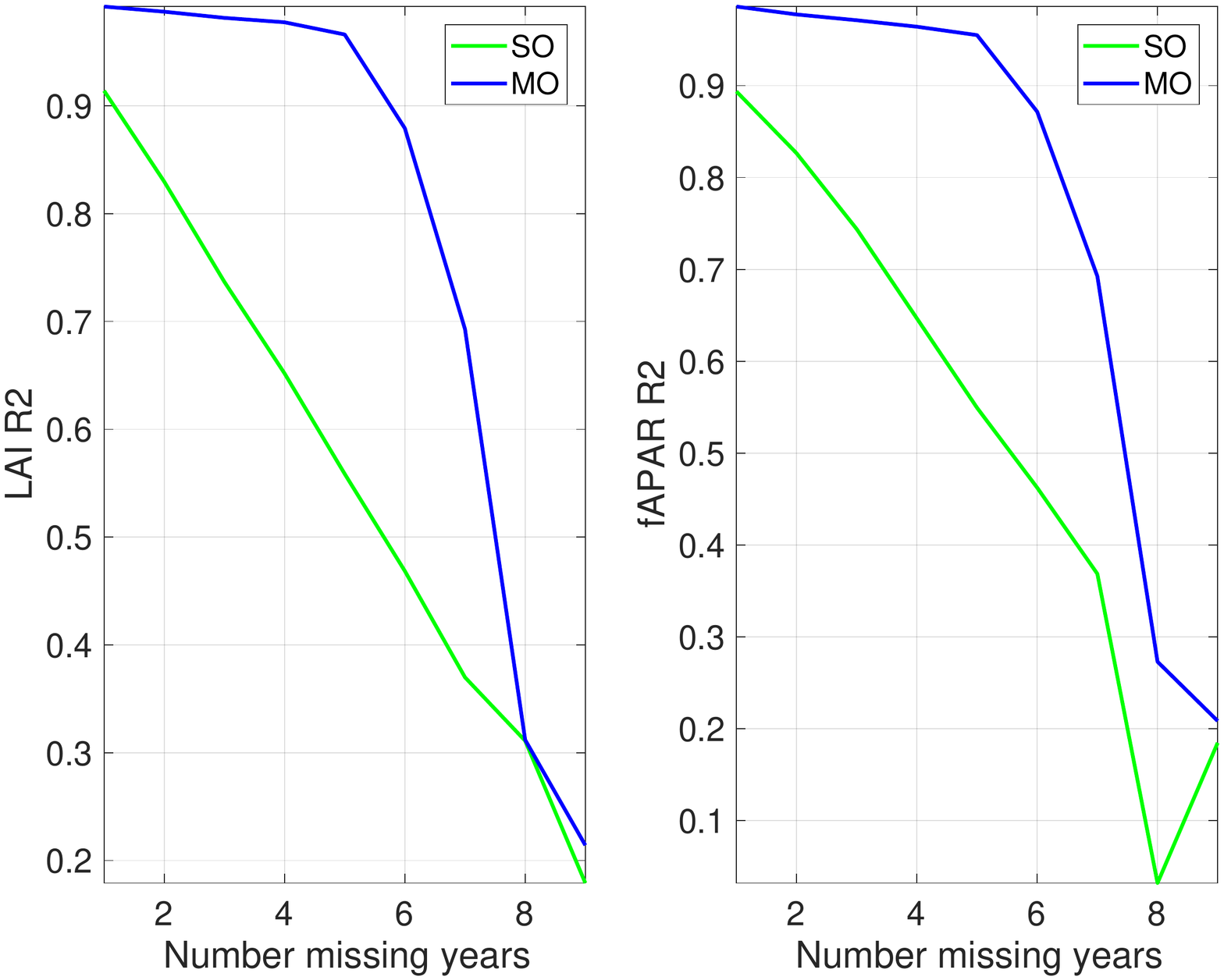}
	\caption{Evolution of the coefficient \(R^2\) for both LAI (left) and fAPAR (right) w.r.t. number of missing years.}
    \label{fig:evolutionR2}
\end{figure}

We repeated the experiments with an increasing number of full missing years. Figure~\ref{fig:evolutionR2} shows how the coefficient \(R^2\) decreases as the number of missing years increases. %
In all tests, the coefficient \(R^2\) for MO is greater than SO for both variables, but for SO it gradually decreases, while for MO it drops more abruptly. The decrease in the error of the MO-GP happens when more than five years are removed, therefore not having a common period in both time series.

\subsection{Capturing time structure}

\begin{figure}[h!] %
	\hspace{-1cm}
	\includegraphics[width= 10cm,height= 4cm]{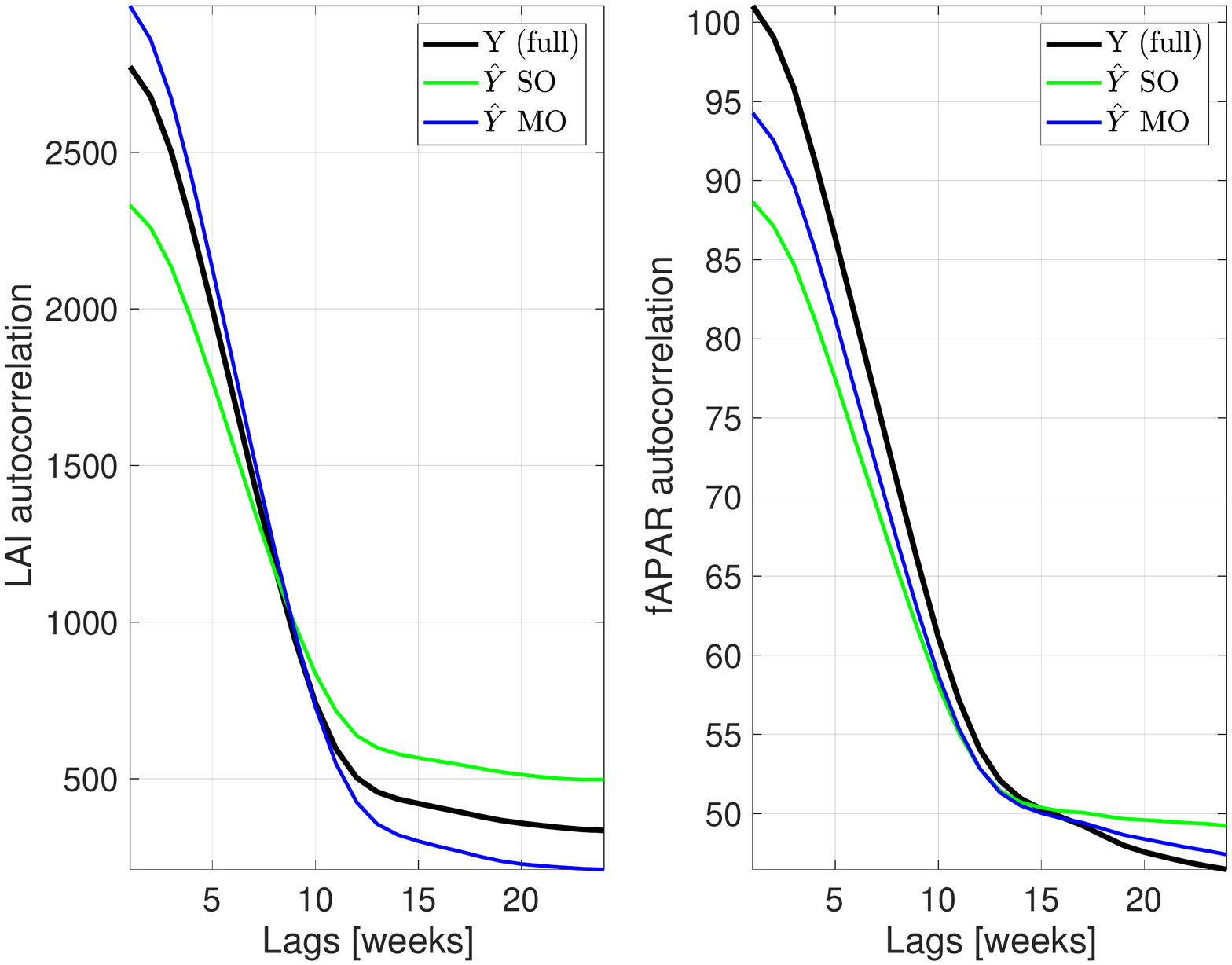}
	\caption{Autocorrelation curves for both LAI (left) and fAPAR (right) w.r.t. number of weeks. This curves show how different models capture the series time structure.}
    \label{fig:autocorrelation}
\end{figure}

In Fig.~\ref{fig:autocorrelation} we show the autocorrelation functions (ACFs) of the actual time series, and their SO and MO predictions. The function gives us a summary about the time structure captured by the models. For both variables, the MO model (blue curve) shows a closer ACF to the actual time series ACF (black curve) than SO does (green curve). %

\section{Conclusions} \label{sec:conclusions}

We have shown in this work how estimating several biophysical parameters simultaneously %
helps to attain consistent predictions and for gap filling scenarios where the missing  information in one variable can be complemented by another one. Multi-output regression is also convenient because only one model is needed, which reduces the computational workload. %

We evaluated a multiple-output GP model based on linear co-regionalization to solve a time series gap filling problem. We used MODIS LAI and fAPAR times series of 11 years. Results showed that the presented model is able to successfully predict both variables simultaneously in regions where no training samples are available by intrinsically exploiting the relationships between the considered output variables, LAI and fAPAR. Future work will consider more sophisticated methods of variable coupling and sparse GPs.

\small
\bibliography{mybib}
\bibliographystyle{IEEEbib}

\end{document}